\documentclass[conference]{IEEEtran}
\IEEEoverridecommandlockouts
\usepackage{cite}
\usepackage{amsmath,amssymb,amsfonts}

\usepackage{algorithmic}
\usepackage{graphicx}
\usepackage{algorithm}
\usepackage{textcomp}
\usepackage{xcolor}
\usepackage{amsmath,amsfonts,amsthm, bm} %
\usepackage{cuted}
\usepackage{multirow}
\usepackage{float} 
\usepackage{booktabs} 
\usepackage{makecell} 

\renewcommand{\baselinestretch}{1.0}

\def\BibTeX{{\rm B\kern-.05em{\sc i\kern-.025em b}\kern-.08em
    T\kern-.1667em\lower.7ex\hbox{E}\kern-.125emX}}
    
\begin{document}

\title{A DAFT Based Unified Waveform Design Framework for High-Mobility Communications\\
	\thanks{This work is funded by Guangdong Natural Science Foundation under Grant 2019A1515011622.   (* tangyq8@mail.sysu.edu.cn)}
}

\author{\IEEEauthorblockN{Xingyao Zhang$^1$, Haoran Yin$^1$, Yanqun Tang*$^1$, Yu Zhou$^1$, Yuqing Liu$^1$, Jinming Du$^1$, Yipeng Ding$^2$}
	\IEEEauthorblockA{
		${}^1$\textit{School of Electronics and Communication Engineering, Sun Yat-sen University, China}
		\\
       ${}^2$\textit{School of Electronic Information, Central South University, China}
      }
}

\maketitle

\begin{abstract}
\textbf{
With the increasing demand for multi-carrier communication in high-mobility scenarios, it is urgent to design new multi-carrier communication waveforms that can resist large delay-Doppler spreads. Various multi-carrier waveforms in the transform domain were proposed for the fast time-varying channels, including orthogonal time frequency space (OTFS), orthogonal chirp division multiplexing (OCDM), and affine frequency division multiplexing (AFDM). Among these, the AFDM is a strong candidate for its low implementation complexity and ability to achieve optimal diversity. This paper unifies the waveforms based on the discrete affine Fourier transform (DAFT) by using the chirp slope factor $k$ in the time-frequency representation to construct a unified design framework for high-mobility communications. The design framework is employed to verify that the bit error rate performance of the DAFT-based waveform can be enhanced when the signal-to-noise ratio (SNR) is sufficiently high by adjusting the chirp slope factor $k$.}
\end{abstract}



\section{Introduction}

Next-generation wireless systems (e.g., Beyond fifth-generation/sixth-generation (B5G/6G)) are anticipated to fulfill the necessities of high throughput, low latency, and extensive communication in high-mobility scenarios, including vehicle-to-everything (V2X) communications, high-speed railways, and unmanned aerial vehicle (UAV) networking. However, due to the heavy Doppler shifts induced in the high-mobility scenarios, the performance of traditional communication waveforms, such as orthogonal frequency division multiplexing (OFDM)\cite{b1}, degrades significantly. Consequently, it is essential to investigate the potential of novel waveforms capable of resisting serious delay-Doppler spreads.

In recent years, designing a new waveform in the transform domain has emerged as an effective way to cope with fast time-varying channels. Communication in the transform domain refers to modulating the information symbols on a set of orthogonal bases through a specific transform. It can be adjusted to the channel characteristics to achieve near-optimal performance. The orthogonal chirp division multiplexing (OCDM)\cite{b2}, which is based on the discrete Fresnel transform (DFnT), modulates signals in the chirp domain and outperforms OFDM in time-dispersive channels thanks to a higher diversity order\cite{b3}. The orthogonal time frequency space (OTFS)\cite{b4}, which employs the Zak transform to modulate a two-dimensional (2D) grid of information symbols directly onto the delay-Doppler domain, also demonstrates superior performance. As a more generalized chirp domain modulation method, affine frequency division multiplexing (AFDM)\cite{b5}\cite{b6}, which employs multiple orthogonal information-bearing chirps generated using the inverse discrete affine Fourier transform (IDAFT)\cite{b7}, can be compatible with OCDM and achieve performance comparable to OTFS. With two adjustable parameters $c_1$ and $c_2$, AFDM is regarded as a strong promising candidate for high-mobility B5G systems.

Given the multitude of B5G/6G application scenarios, it is a major challenge to develop novel waveforms with robust compatibility. The typical solutions can be divided into two categories, one is altering basis functions for different transform domains, and the other is changing the number of parameters to generalize the transform type. The orthogonal time chirp space modulation based on fractional Fourier transform (OTCS-FrFT)\cite{b8} combines the advantages of linear chirps and OTFS modulation, introducing signals into the delay-fractional Doppler domain with FrFT. Moreover, the chirp convolved data transmission (CCDT)\cite{b9} employs a second-order polynomial to define the phase of the exponential function as the chirp basis function. Without the DFT/IDFT operations, the three polynomial coefficients of CCDT can be selected to adapt the waveform to the channel while simultaneously producing orthogonal signals. Furthermore, the offset linear canonical transform (OLCT)\cite{b10}, as a variant of linear canonical transform (LCT), is capable of accommodating a variety of basis functions with two additional parameters to LCT, which means it has six parameters to cover a greater range of transforms. However, the above specific methods present a challenge in achieving a great balance between the number of adjustable parameters and the implementation complexity.

This paper delves into the multi-carrier waveform design architecture in the transform domain based on discrete affine Fourier transform (DAFT). In consideration of the compatibility of 5G, AFDM with two adjustable parameters is particularly suitable for the delay-Doppler domain, as it allows $c_1=c_2=0$ to be compatible with OFDM\cite{b5,b6}. Developed from FrFT\cite{b11}, AFDM takes DAFT as the core transform, which can be conducted quickly by two chirp multiplications and one FFT operation. Furthermore, DAFT has one more degree of freedom than FrFT. We propose a method based on DAFT to unify three widely-used waveforms, namely OFDM, OCDM, and AFDM by using the chirp slope factor $k$ in the time-frequency representation. Using this method, we can see the distribution of subcarriers on the time-frequency plane during the chirp domain modulation. In addition, it is possible to select the parameters of AFDM reasonably to obtain OFDM and OCDM, as well as to enhance the bit error rate (BER) performance of AFDM.

In this paper, after a brief overview of OFDM, OCDM, and AFDM, we compare the BER of these three waveforms under different conditions. Then we construct a DAFT-based unified waveform design framework for for high-mobility communications. In particular, the parameter selection strategy is investigated, and the BER curve for the optimal parameter selection is obtained. The simulation results demonstrate the feasibility of the proposed unified framework, and the optimized AFDM waveform can achieve better BER performance than the conventional AFDM when the signal-to-noise ratio (SNR) is sufficiently high. The paper is organized as follows. The OCDM and AFDM are introduced in Section II. A unified waveform design scheme is established in Section III. Simulation results are provided in Section IV, while Section V concludes the paper.

$Notations:$ This paper will employ the following notations. The symbols $a$, $\mathbf{a}$, and $\mathbf{A}$ represent a scalar, vector, and matrix, respectively. $(\cdot)^T$ is the transpose operation, $(\cdot)^H$ is the Hermitian transpose operation, and $\mathbf{I}$ denotes the unit matrix.

\section{ Basic Concepts of OCDM and AFDM}
This section reviews the modulation and demodulation processes of OCDM and AFDM\cite{b2,b5,b6}. Assume that the wireless channel is time-varying and has the following impulse response at time $n$ and delay $l$:
\begin{equation}
	g_{n}(l)=\sum_{i=1}^{P} h_{i} e^{-j 2 \pi \frac{\alpha_{i}}{N} n} \delta\left(l-l_{i}\right),
\end{equation}
where $P$ is the number of paths, $h_{i}$ is the complex gain, $l_{i}$ and $\alpha_{i}$ are the delay and Doppler normalized with sample period and subcarrier spacing, respectively. 

\subsection{OCDM}
Let $\boldsymbol{x}$ denote the $N \times 1$ vector of quadrature amplitude modulation (QAM) symbols. After the serial to parallel operation, $N$-points inverse DFnT (IDFnT) is performed to map $\boldsymbol{x}$ to the time domain as\cite{b2}
\begin{equation}
    s_n=\frac{1}{\sqrt{N}} e^{j \frac{\pi}{4}} \sum_{m=0}^{N-1} x_m \times\left\{\begin{array}{l}e^{-j \frac{\pi}{N}(n-m)^2} \quad N \equiv 0(\bmod 2) \\ e^{-j \frac{\pi}{N}\left(n-m-\frac{1}{2}\right)^2} N \equiv 1(\bmod 2) .\end{array}\right.
\end{equation}

The DFnT matrix can be decomposed using the equation $\boldsymbol{\Phi}=\boldsymbol{\Theta}_2 \mathbf{F} \boldsymbol{\Theta}_1$, where $\mathbf{F}$ is the DFT matrix with entries $e^{-j 2 \pi m n / N} / \sqrt{N}$, $\boldsymbol{\Theta}_1$ and $\boldsymbol{\Theta}_2$ are two diagonal matrices given by
\begin{equation}
    \Theta_1(m)=e^{-j \frac{\pi}{4}} \times\left\{\begin{array}{ll}e^{j\frac{\pi}{N} m^2} & N \equiv 0(\bmod 2) \\e^{j \frac{\pi}{4N}} e^{j \frac{\pi}{N}}\left(m^2+m\right) & N \equiv 1(\bmod 2)\end{array} \right.
\end{equation}
and
\begin{equation}
     \Theta_2(n)=\left\{\begin{array}{ll}e^{j \frac{\pi}{N} n^2} & N \equiv 0(\bmod 2) \\e^{j \frac{\pi}{N}\left(n^2-n\right)} & N\equiv 1(\bmod 2).
    \end{array} \right.
\end{equation}
(2) can be rewritten in matrix form as
\begin{equation}
    \mathbf{s}=\boldsymbol{\Theta}_1{ }^H \mathbf{F}^H \boldsymbol{\Theta}_2{ }^H \boldsymbol{x}=\boldsymbol{\Phi}^H \boldsymbol{x} .
\end{equation}

The received signal $r_n$ after going through the delay-Doppler channel in (1) is given by
\begin{equation}
    r_n=\sum_{l=0}^{\infty} s_{n-l} g_n(l)+w_n ,
\end{equation}
then $r_n$ can be rewritten in matrix form as 
\begin{equation}
    \mathbf{r}=\mathbf{H} \mathbf{s}+\boldsymbol{w}=\mathbf{H} \boldsymbol{\Phi}^H \boldsymbol{x}+\boldsymbol{w}.
\end{equation}

After discarding the cyclic prefix (CP) and demodulation with $N$-points DFnT, the received chirp domain samples can be presented as
\begin{equation}
    \boldsymbol{y}=\boldsymbol{\Theta}_2 \mathbf{F} \boldsymbol{\Theta}_1 \mathbf{H} \boldsymbol{\Theta}_1{ }^H \mathbf{F}^H \boldsymbol{\Theta}_2{ }^H \boldsymbol{x}+\tilde{\boldsymbol{w}}=\mathbf{H}_{\mathrm{eff}} \boldsymbol{x}+\tilde{\boldsymbol{w}},
\end{equation}
where $\tilde{\boldsymbol{w}} \sim \mathcal{C N}\left(\mathbf{0}, N_0 \mathbf{I}\right)$ is an additive Gaussian noise vector with variance $N_0$ and $\mathbf{H}$ is the channel matrix in the time domain.

\subsection{AFDM}
Let $\boldsymbol{x}$ denote the $N \times 1$ vector of QAM symbols in the DAFT domain. The $N$-points IDAFT is performed to map $\boldsymbol{x}$ to the time domain as\cite{b5}
\begin{equation}
s_{n}=\frac{1}{\sqrt{N}} \sum_{m=0}^{N-1} x_{m} e^{j 2 \pi\left(c_{2} m^{2}+\frac{1}{N} m n+c_{1} n^{2}\right)},
\end{equation}
with $n=0, \cdots, N-1$, where $N$ denotes the number of subcarriers, $c_1$ and $c_2$ are two fundamental parameters of AFDM. Before transmitting $\mathbf{s}$ into the channel, a $chirp-periodic$ prefix (CPP) should be added with a length of $L_{\mathrm{cpp}}$, which is any integer greater than or equal to the value in samples of the maximum delay spread of the channel. The prefix is
\begin{equation}
   s_n=s_{N+n} e^{-j 2 \pi c_1\left(N^2+2 N n\right)}, \quad n=-L_{\mathrm{cpp}}, \cdots,-1 . 
\end{equation}

Similar to OCDM, the CPP is removed after passing through the channel and DAFT is performed on the received signal $r_n$ to obtain the receiver samples $y_m$. The DAFT domain output symbols are obtained by
\begin{equation}
    y_m=\frac{1}{\sqrt{N}} \sum_{n=0}^{N-1} r_n e^{-j 2 \pi\left(c_2 m^2+\frac{1}{N} m n+c_2 n^2\right)}.
\end{equation}

Hence, the relationship between $s_n$ and $x_m$ can be written as
\begin{equation}
   \mathbf{s}=\boldsymbol{\Lambda}_{c_1}^H \mathbf{F}^H \boldsymbol{\Lambda}_{c_2}^H \boldsymbol{x},
\end{equation}
where
\begin{equation}
    \boldsymbol{\Lambda}_{c_i}=\operatorname{diag}\left(e^{-j 2 \pi c_i n^2}, n=0,1, \ldots, N-1\right) .
\end{equation}

The received sample matrix can be presented as
\begin{equation}
     \begin{aligned}
     \boldsymbol{y} & =\boldsymbol{\Lambda}_{c_2} \mathbf{F}\boldsymbol{\Lambda}_{c_1} \mathbf{r} \\
     & =\sum_{i=1}^P h_i \boldsymbol{\Lambda}_{c_2} \mathbf{F} \boldsymbol{\Lambda}_{c_1} \boldsymbol{\Gamma}_{\mathrm{CPP}_i} \boldsymbol{\Delta}_{f_i} \boldsymbol{\Pi}^{l_i} \boldsymbol{\Lambda}_{c_1}^H \mathbf{F}^H \boldsymbol{\Lambda}_{c_2}^H \boldsymbol{x}+\tilde{\boldsymbol{w}}\\
     & =\sum_{i=1}^P h_i\mathbf{H}_i\boldsymbol{x}+\tilde{\boldsymbol{w}}\\
     & =\mathbf{H}_{\mathrm{eff}} \boldsymbol{x}+\tilde{\boldsymbol{w}},
      \end{aligned}
\end{equation}
where $f_i$ is the Doppler shift (in digital frequencies) ,  $\Delta_{f_i}\triangleq\operatorname{diag}\left(e^{-j 2 \pi f_i n}, n=0,1, \ldots, N-1\right)$ and $\mathbf{\Pi}$ is the forward cyclic-shift matrix which can be written as
\begin{equation}
\boldsymbol{\Pi}=\left[\begin{array}{cccc}
0 & \cdots & 0 & 1 \\
1 & \cdots & 0 & 0 \\
\vdots & \ddots & \ddots & \vdots \\
0 & \cdots & 1 & 0
\end{array}\right]_{N \times N}.
\end{equation}
$\boldsymbol{\Gamma}_{\mathrm{CPP}_i}$ is a diagonal array of $N\times N$ and can be written as
\begin{equation}
   \begin{aligned}
& \boldsymbol{\Gamma}_{\mathrm{CPP}_i} \triangleq \\
& \quad \operatorname{diag}\left(\left\{\begin{array}{ll}
e^{-j 2 \pi c_1\left(N^2-2 N\left(l_i-n\right)\right)} & n<l_i \\
1 & n \geq l_i
\end{array}\right)\right.,
\end{aligned}
\end{equation}
the effective channel matrix $\mathbf{H}_{\mathrm{eff}} \triangleq \boldsymbol{\Lambda}_{c_2} \mathbf{F} \boldsymbol{\Lambda}_{c_1} \mathbf{H} \boldsymbol{\Lambda}_{c_1}^H \mathbf{F}^H \boldsymbol{\Lambda}_{c_2}^H$, $\tilde{\boldsymbol{w}}=\boldsymbol{\Lambda}_{c_2} \mathbf{F} \boldsymbol{\Lambda}_{c_1}\boldsymbol{w}$ and $\boldsymbol{w} \sim$ $\mathcal{C N}\left(\mathbf{0}, N_0 \mathbf{I}\right)$. Since $\boldsymbol{\Lambda}_{c_2} \mathbf{F} \boldsymbol{\Lambda}_{c_1}$ is a unitary matrix, $\tilde{\boldsymbol{w}}$ and $\boldsymbol{w}$ have the same covariance.

The structure of the DAFT and DFnT matrices is strikingly similar, with both matrices containing the DFT process. It can be readily demonstrated that the DFT and DFnT can be obtained by setting $c_1=c_2=0$ and $c_1=c_2=\frac{1}{2N}$ in the DAFT, respectively. This implies that the AFDM waveform can be compatible with the other two.

\section{Unified Waveform Design Framework}
\subsection{Elements of Waveform Design}
\begin{figure*}[htbp]
	\centering
\includegraphics[width=0.86\textwidth,height=0.36\textwidth]{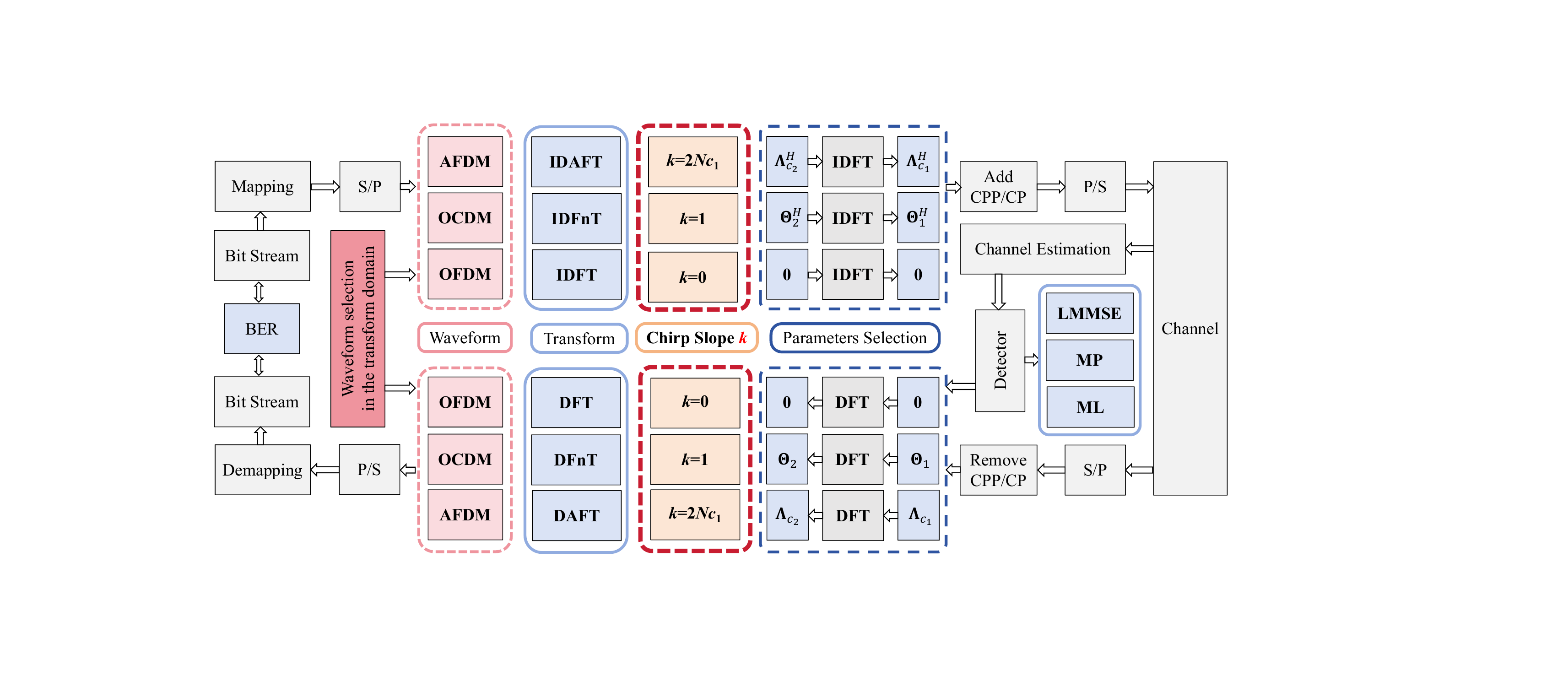}
	\caption{Schematic diagram of the proposed unified Waveform design framework.}
	\label{4-1}
\end{figure*}

The waveform design framework involves various elements such as constellation mapping, waveform transform, cyclic prefix alternation, serial-parallel conversion, channel modeling, estimation, and detection as Fig. \ref{4-1}. These steps typically occur symmetrically, demanding consideration of both forward and backward transforms to minimize BER. The fundamental aspect of waveform design lies in selecting the appropriate core transform for data modulation, accompanied by the incorporation of delay-Doppler channel modeling to simulate diverse scenarios.  In the following, we focus on OFDM, OCDM, and AFDM for our study, which can be easily linked to each other by the parameters of DAFT. The unified design framework for these three waveforms is shown in Fig. \ref{4-1}, which contains the fundamental elements of waveform design.

\subsection{Principles of Waveform Design}
In OFDM, we conduct the modulation by performing IDFT at the transmitter and demodulation via DFT at the receiver. Whereas in OCDM, due to the chirp domain modulation, we can achieve better performance by multiplying the chirp phase matrix $\boldsymbol{\Theta}_1$ and $\boldsymbol{\Theta}_2$ factors before and after the DFT matrix during digital implementation as shown in (5). Nevertheless, it should be noted that these two waveforms do not achieve full diversity in doubly dispersive channels. In AFDM, we can set $c_1$ and $c_2$ so that the transform domain impulse response constitutes a full delay-Doppler representation of the channel.

Then it is essential to ensure that the delay components and Doppler components in the equivalent channel matrix do not overlap with one another, to guarantee that each path is effectively separated. As illustrated in \cite{b5}, a comprehensive delay-Doppler representation within the DAFT domain can be attained when the condition 
\begin{equation}
    2 \alpha_{\max } l_{\max }+2 \alpha_{\max }+l_{\max }<\mathrm{N}
\end{equation}
is met, with $c_1$ satisfy
\begin{equation}
   c_1=\frac{2 \alpha_{\max }+1}{2 N}
\end{equation}
and $c_2$ is set to be an arbitrary irrational number or a rational number sufficiently smaller than $1/(2N)$.

Although (18) provides the principle related to the optimal selection of $c_1$, it should be noted that the parameters of AFDM are not fixed. It is possible to take a certain range under the principle of AFDM parameter selection and obtain the most suitable parameters by simulating. 

\subsection{Unified Waveform Design Methodology}
\begin{figure*}[htbp]
	\centering
\includegraphics[width=0.9\textwidth,height=0.22\textwidth]{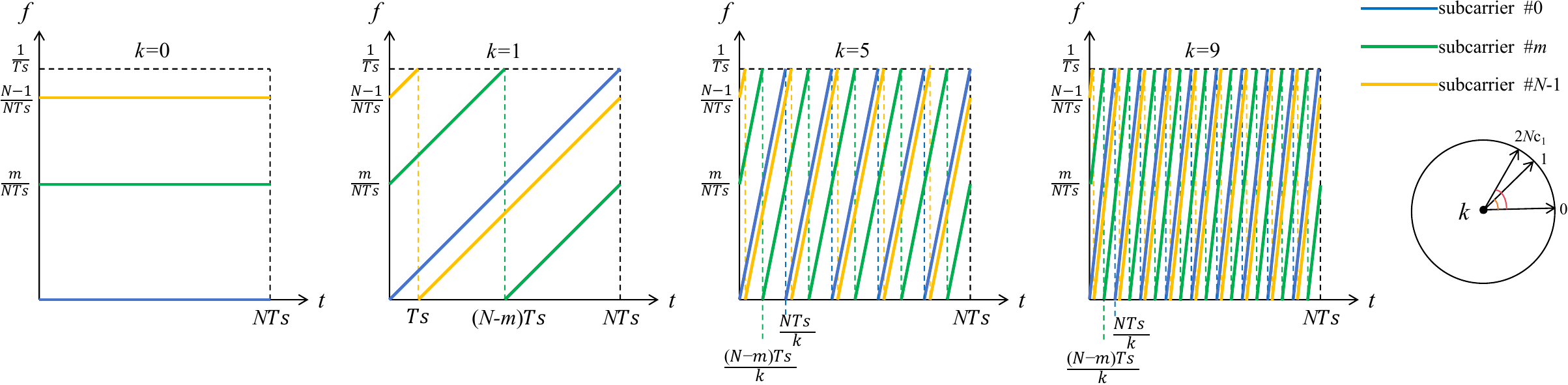}
	\caption{Schematic diagram of the proposed unified waveform design methodology. ($T_s$ is the sampling period and $k$ is the chirp slope.)}
	\label{4-3}
\end{figure*}

Given that both OCDM and AFDM can be understood as communication waveforms in the chirp domain while OFDM can also be understood as a special kind of chirp modulation with the chirp rate of zero, we initially draw the subcarrier distributions of OFDM, OCDM, and AFDM in the time-frequency plane based on different forms of DAFT. As shown in Fig. \ref{4-3}, the blue line represents the $0$th subcarrier, the green line denotes the $m$th subcarrier and the yellow line indicates the $(N-1)$st subcarrier. The three waveforms are unified by applying the chirp slope $k$ of the subcarriers in the time-frequency plane. When $k$ is set to 0, each subcarrier occupies a small segment of a specific frequency spectrum to obtain OFDM. When $k$ is set to 1, the result is OCDM, and the subcarrier slices the time-frequency plane resources at a 45-degree angle. Finally, when $k$ is set to $2Nc_1$, the result is AFDM, which also corresponds with the parameters of $c_1$ in AFDM.

\subsection{Waveform Design with Different Parameter Choices}
Table \ref{t1} summarizes the transform relationship presented in the unified waveform design framework. It is evident from this table that the resource distribution of the waveform in the time-frequency plane is only related to the chirp slope $k$ of the chirp modulation. This implies that the tuning of the parameter $c_1$ is crucial in AFDM.

Since heuristic search methods have been proposed for coefficient selection for more elaborate channel models\cite{b9}, we restrict the search to the interval [0, 1] and identify all feasible values of $c_1$ and $c_2$. Once the feasible values for the parameters are determined, further exploration of suitable values within smaller intervals may be conducted by increasing the range or reducing the step size appropriately. Additional details regarding the simulation methodology can be found in Section IV.

\begin{table}[]
	\renewcommand\arraystretch{1.7}
    \centering
	\caption{Waveform Parameter Selection Table}
	\label{t1}
    \begin{tabular}{cccc}
    \toprule \multirow{2}{*}{ Waveform } & & Parameter Selection & \\
     & $c_1$ & $c_2$ & $k$ \\
    \midrule OFDM & 0 & 0 & 0 \\
     OCDM & $\frac{1}{2N}$ & $\frac{1}{2N}$ & 1 \\
     AFDM & $\frac{2\alpha_{\max}+1}{2N}$ & \makecell[c]{smaller than $\frac{1}{2N}$ \\or ${\forall}$ irrational number}  & $2Nc_1$ \\
    \bottomrule
    \end{tabular}
\end{table}

\section{Simulation Results}
In this section, we first analyze the performance in terms of BER of OFDM, OCDM, and AFDM with different channel conditions and simulation settings. We compare the performance of the three waveforms with ideal CSI in four main aspects: different modulation modes, different numbers of paths,  different speeds, and different detectors. In addition, the OTFS is also simulated for comparison. We choose $N=256$ for OFDM, OCDM, and AFDM, and  $N_{\mathrm{OTFS}}=16$, $M_{\mathrm{OTFS}}=16$ for OTFS to ensure all the considered waveforms occupy the same resources. Then, we determine the optimal parameter selection for the AFDM waveforms in a variable-step search manner. After that, we implement a performance comparison of OFDM, OCDM, and AFDM by choosing the time-frequency factor $k$ proposed in section III. 

\begin{table}[]
    \renewcommand\arraystretch{1.3}
    \centering
    \caption{Simulation Parameters}
    \begin{tabular}{|c|c|c|}
    \hline
        \textbf{Parameter} & \textbf{Symbol} & \textbf{Value} \\
    \hline
       Carrier frequency  & $f_0$ & 4GHz \\
       \hline
       Subcarrier spacing  & $\Delta f$ & 1kHz \\
       \hline
       Number of chirps  & $N$ & 256 \\
       \hline
       Modulation  & - & 4QAM \\
       \hline
       Channel estimation  & - & ideal \\
       \hline
       Number of paths  & $P$ & 3 \\
       \hline
       Normalized delay & $l_{\max}$ & 2 \\
       \hline
       Normalized Doppler shift &  $\alpha_{\max}$ & 2 \\
       \hline
       Maximum Doppler shift & $f_{\max}$ & 2kHz \\
       \hline
       Maximum speed & $v_{\max}$ & 540km/h\\
       \hline
       Detector  & - & LMMSE \\
    \hline     
    \end{tabular}
    \label{t2}
\end{table}

The base simulation parameters for all waveforms are shown in Table \ref{t2}. In all simulations, the complex gains $h_i$ are generated as independent complex Gaussian random variables with zero mean and $1/P$ variance, the delay shifts are fixed and Jakes Doppler spectrum is considered for each channel realization, i.e, the Doppler shifts are varying and the Doppler shift of each path is generated using $\alpha_i=\alpha_{\max } \cos \left(\theta_i\right)$, where $\theta_i$ is uniformly distributed over $[-\pi, \pi]$. 

\begin{figure}[htbp]
	\centering
\includegraphics[width=0.32\textwidth,height=0.25\textwidth]{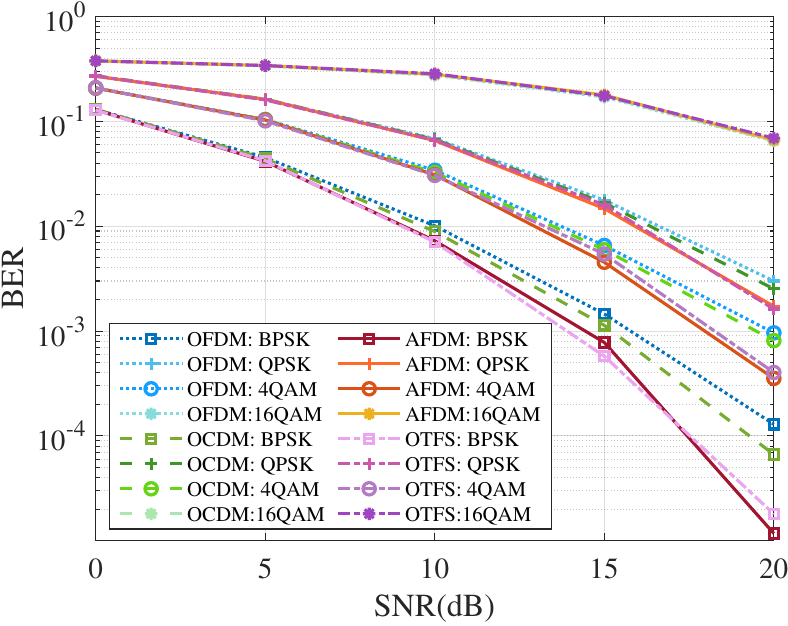}
	\caption{BER performance of OFDM, OCDM, AFDM, and OTFS with $N = 256$, $N_{\mathrm{OTFS}}=16$, $M_{\mathrm{OTFS}}=16$ using different modulation modes and LMMSE detector assuming a 3-path LTV channel with $l_{\max}$ = 2 and $\alpha_{\max}$ = 2}
	\label{3-7}
\end{figure}

Fig. \ref{3-7} illustrates the BER performance curves of the three waveforms and OTFS under different modulation modes, mainly comparing the four modulation modes of BPSK, QPSK, 4QAM, and 16QAM. It can be demonstrated that AFDM exhibits superior performance to that of OCDM and OFDM in each modulation mode. Furthermore, AFDM and OTFS exhibit similar BER performances.

\begin{figure}[htbp]
	\centering
\includegraphics[width=0.32\textwidth,height=0.25\textwidth]{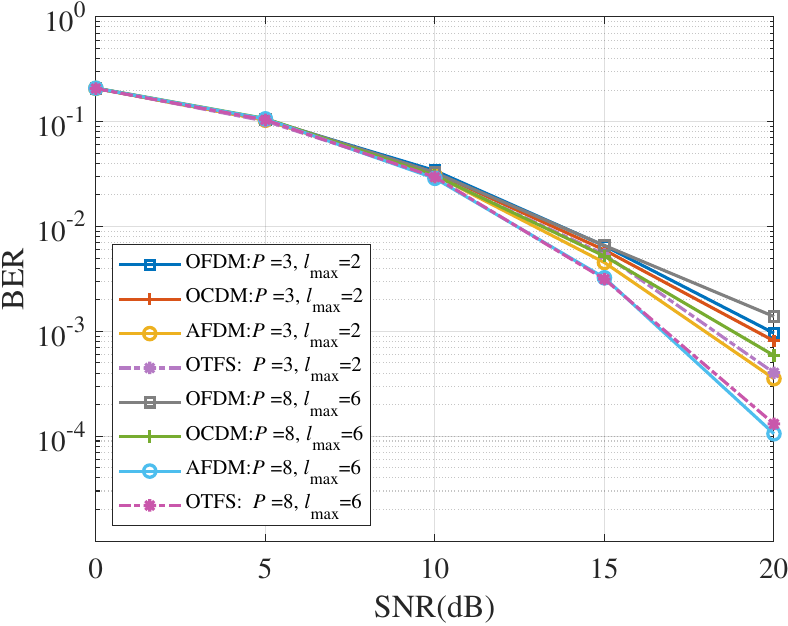}
	\caption{BER performance of OFDM, OCDM, AFDM, and OTFS with $N = 256$, $N_{\mathrm{OTFS}}=16$, $M_{\mathrm{OTFS}}=16$ using 4QAM and LMMSE detector in different number of paths with $\alpha_{\max}$ = 2 and different $l_{\max}$}
	\label{3-8}
\end{figure}

Fig. \ref{3-8} compares the BER performance of the three waveforms and OTFS when the number of paths is set to 3 and 8, respectively. In the case of the 3-path configuration, the normalized delay is set to [0, 1, 2], while in the 8-path scenario, it is set to [0, 0, 1, 2, 3, 4, 5, 6]. The comparison indicates that AFDM waveforms exhibit the most robust performance against multipath interference, with a marginal advantage over OTFS. As the probability of path separation increases, the benefits of AFDM waveform design based on DAFT become increasingly evident.

\begin{figure}[htbp]
	\centering
\includegraphics[width=0.32\textwidth,height=0.25\textwidth]{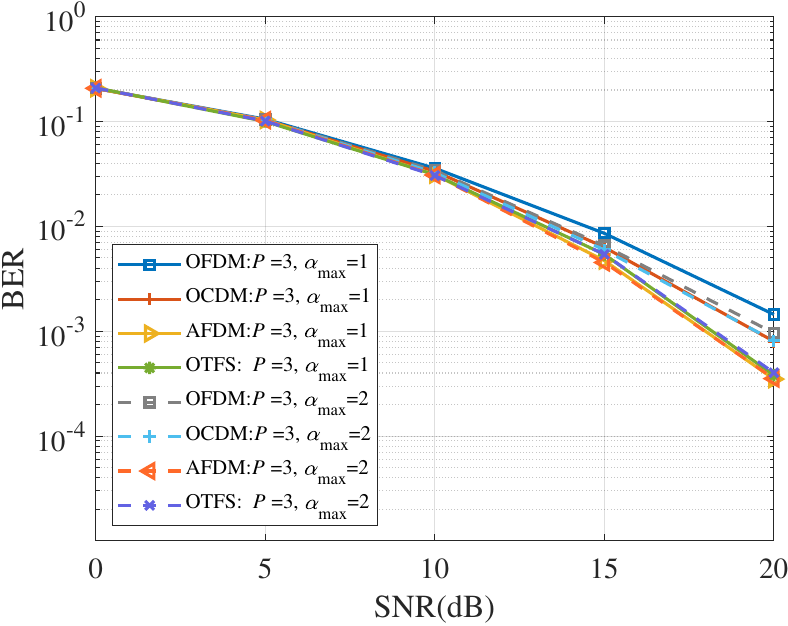}
	\caption{BER performance of OFDM, OCDM, AFDM, and OTFS with $N = 256$, $N_{\mathrm{OTFS}}=16$, $M_{\mathrm{OTFS}}=16$ using 4QAM and LMMSE detector assuming a 3-path LTV channel with different maximum Doppler shift $\alpha_{\max}$.}
	\label{3-10}
\end{figure}

Fig. \ref{3-10} shows the impact of varying speeds of moving objects on the three communication waveforms and OTFS. We set the normalized maximum Doppler shift to 1 and 2, corresponding to speeds of 270km/h and 540km/h, respectively, to compare the BER performance of the waveforms. It can be observed that for the transform domain waveforms including AFDM, OCDM, and OTFS, the BER performance is almost identical at different speeds, which indicates that the transform domain waveforms are robust to Doppler shifts and can be effectively adapted to high-speed mobile scenarios.

\begin{figure}[htbp]
	\centering
\includegraphics[width=0.32\textwidth,height=0.25\textwidth]{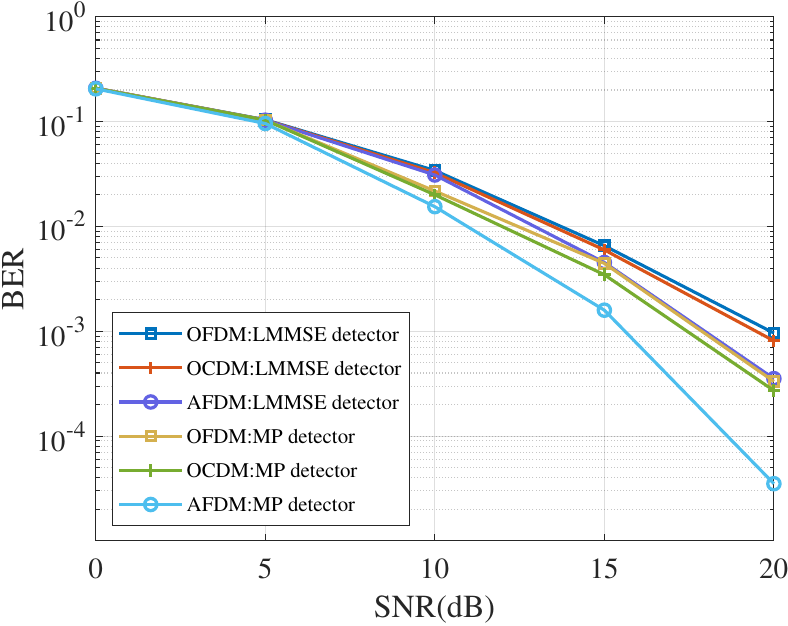}
	\caption{BER performance of OFDM, OCDM, and AFDM with $N = 256$ using 4QAM assuming a 3-path LTV channel with $l_{\max}$ = 2 and $\alpha_{\max}$ = 2 for different detectors.}
	\label{3-11}
\end{figure}

Fig. \ref{3-11} compares the performance of the three waveforms under the linear minimum mean square error (LMMSE) detector and the message passing (MP) detector respectively. It can be observed that the performance of the MP detector is significantly superior to that of LMMSE. This superiority is determined by the advantage of the symbol-by-symbol detection of MP, which makes full use of the mean and variance of the noise.

\begin{figure}[htbp]
	\centering
\includegraphics[width=0.4\textwidth,height=0.3\textwidth]{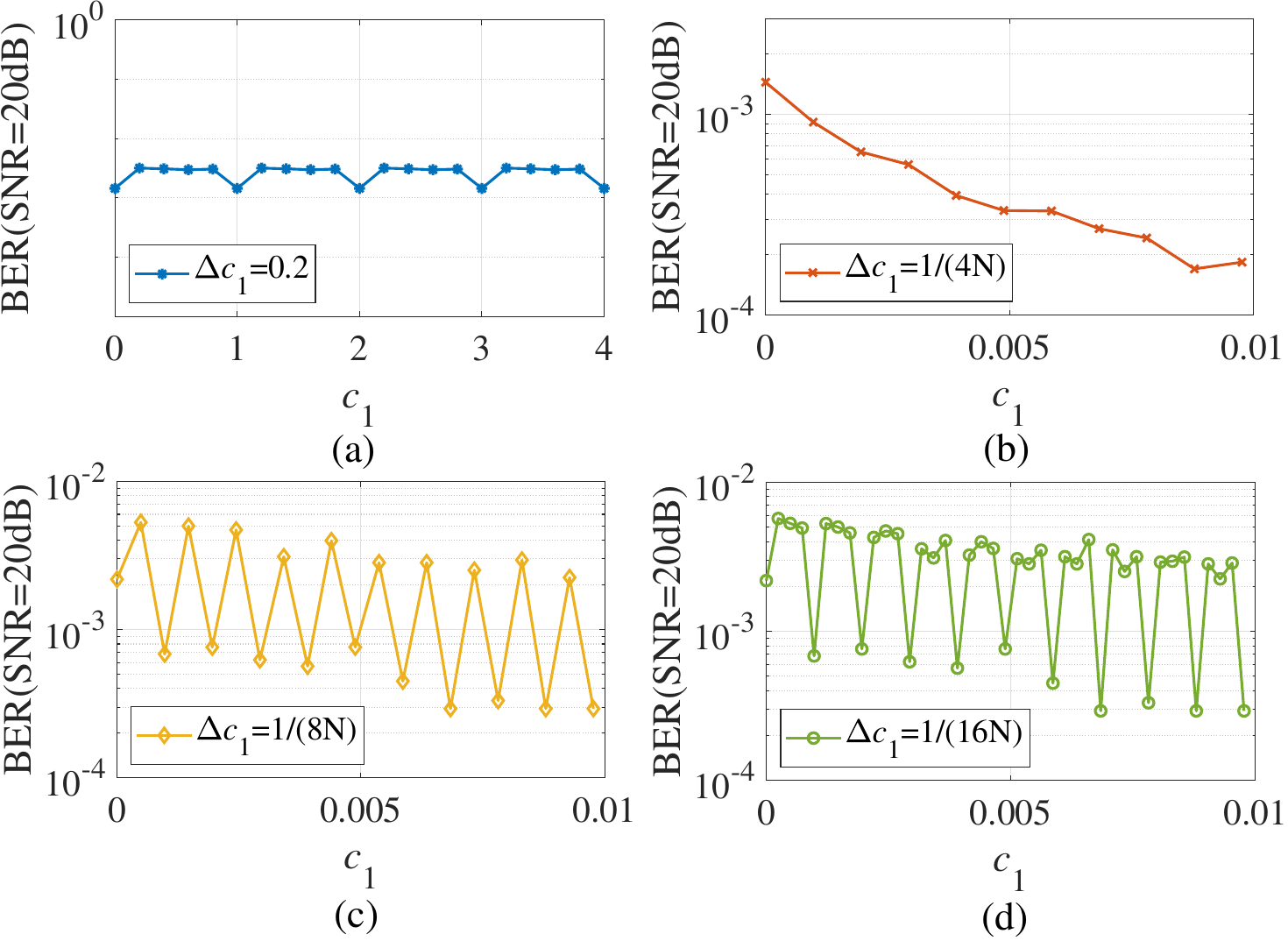}
	\caption{BER performance corresponding to different steps of $c_1$ at SNR\\=20dB. (a)$\Delta$$c_1$=0.2; (b)$\Delta$$c_1$=$\frac{1}{4N}$; (c)$\Delta$$c_1$=$\frac{1}{8N}$; (d)$\Delta$$c_1$=$\frac{1}{16N}$.}
	\label{4-4}
\end{figure}

\begin{figure}[htbp]
	\centering
\includegraphics[width=0.22\textwidth,height=0.15\textwidth]{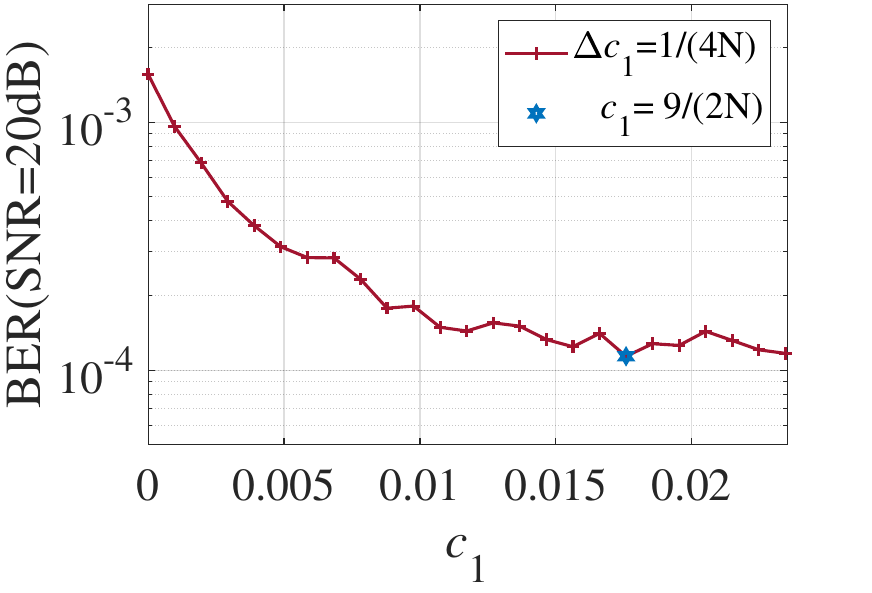}
	\caption{BER performance corresponding to different $c_1$ ($\Delta$$c_1$=$\frac{1}{4N}$) at SNR=20dB.}
	\label{4-5}
\end{figure}

\begin{figure}[htbp]
	\centering
\includegraphics[width=0.45\textwidth,height=0.16\textwidth]{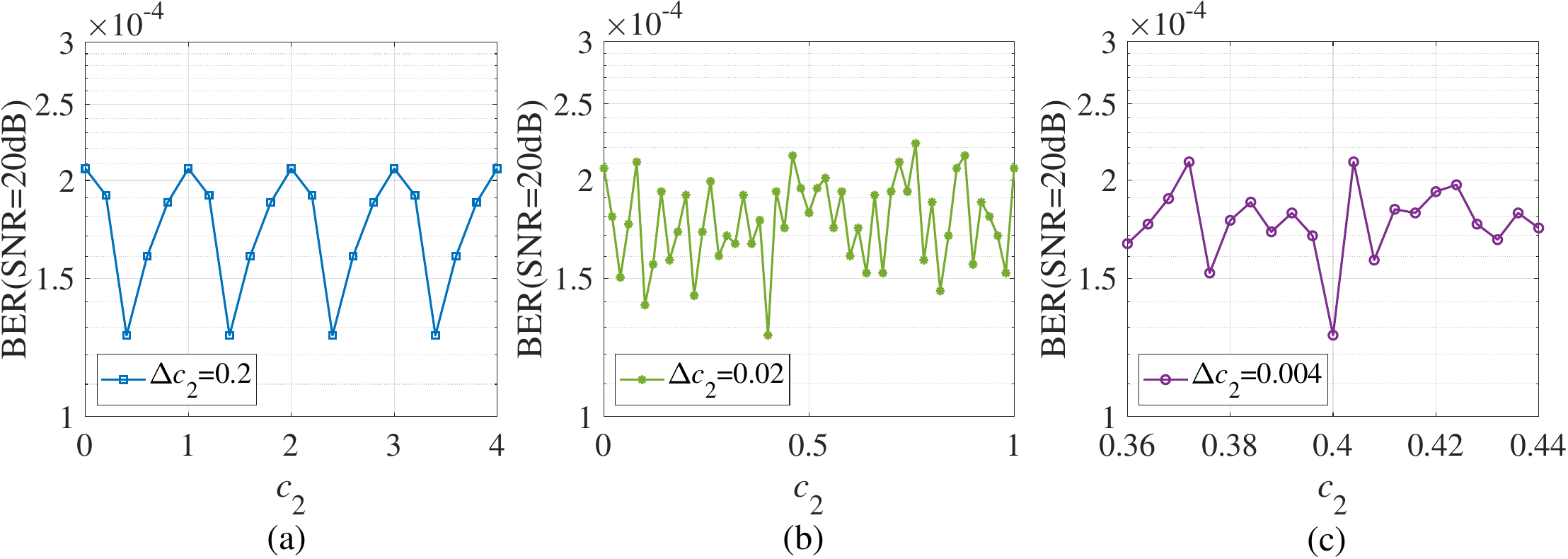}
	\caption{BER performance corresponding to different $c_2$ at SNR=20dB. (a)$\Delta$$c_2$=0.2; (b)$\Delta$$c_2$=0.02; (c)$\Delta$$c_2$=0.004.}
	\label{4-7}
\end{figure}

To find the best parameter selection scheme for BER performance under 8-path channels, we simulate the values of $c_1$ in steps of $\frac{1}{4N}$, $\frac{1}{8N}$, and $\frac{1}{16N}$, respectively. As shown in Fig. \ref{4-4} and Fig. \ref {4-5}, our results indicate that there exists a minimal value of BER when $c_1$ obtains $\frac{9}{2N}$. Similarly, simulating the values of $c_2$, we find that $c_2$ has poor regularity for the AFDM waveform performance. Nevertheless, we find that the best BER performance exists when $c_2 = 0.4$ for our chosen channel as Fig. \ref{4-7}.

\begin{figure}[htbp]
	\centering
\includegraphics[width=0.32\textwidth,height=0.25\textwidth]{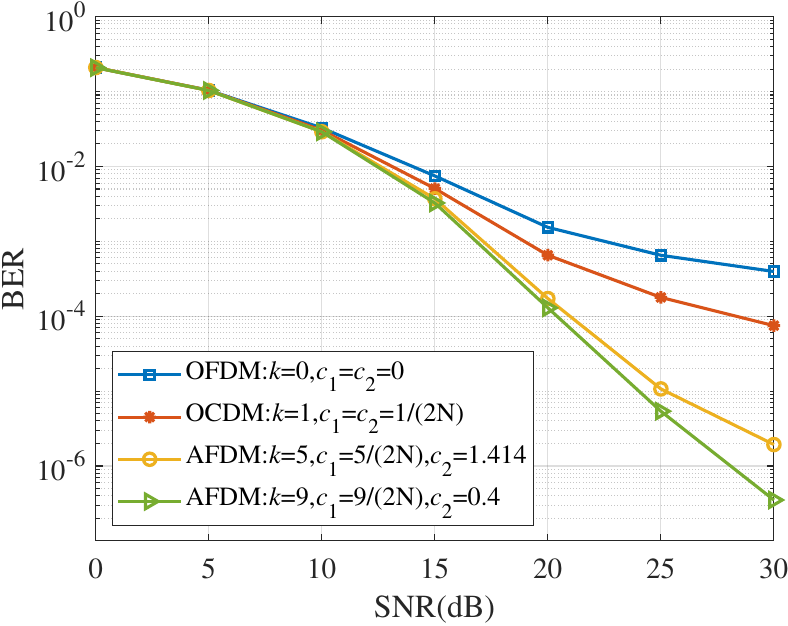}
	\caption{BER versus SNR under different parameter selection schemes.}
	\label{4-8}
\end{figure}
Finally, Fig. \ref{4-8} demonstrates the performance comparison of OFDM, OCDM, and AFDM with two different parameter selection schemes. The results indicate that the concept of designing new waveforms by selecting appropriate parameters is a viable approach. In particular, under conditions of high SNR, the performance of parameter-optimized waveforms can be significantly enhanced.
\section{Conclusion}
In this paper, we propose a DAFT-based waveform design framework that unifies OFDM, OCDM, and AFDM with the help of the chirp slope factor $k$ to distinguish subcarriers from different waveforms in the time-frequency representation. Guided by this framework, we compare the BER performance of the three waveforms under various channel conditions and modulation settings and investigate the potential for optimizing the parameter selection methods to enhance the overall performance of the waveforms. The simulation results show that by choosing the chirp's slope $k$ appropriately, we can improve the BER performance of AFDM under high SNR conditions to resist large delay-Doppler spreads.

\end{document}